\documentclass[lettersize,journal]{IEEEtran}
\usepackage{amsmath,amsfonts}
\usepackage{algorithmic}
\usepackage{algorithm}
\usepackage{array}
\usepackage[caption=false,font=normalsize,labelfont=sf,textfont=sf]{subfig}
\usepackage{textcomp}
\usepackage{stfloats}
\usepackage{url}
\usepackage{verbatim}
\usepackage{graphicx}
\usepackage{cite}
\makeatletter
\let\NAT@parse\undefined
\makeatother
\usepackage{hyperref}
\usepackage{tabularray}
\usepackage{multirow}
\usepackage{colortbl}
\usepackage{color}
\usepackage{CJKutf8}    
\usepackage{booktabs}
\usepackage{arydshln}
\usepackage{pifont}

\usepackage{orcidlink} 

\begin{document}

\title{Factual Serialization Enhancement: A Key Innovation for Chest X-ray Report Generation}

\author{Kang Liu $^{\orcidlink{0000-0002-3150-0185}}$, Zhuoqi Ma $^{\orcidlink{0000-0003-0729-9706}}$, Mengmeng Liu $^{\orcidlink{0000-0002-8740-0386}}$, Zhicheng Jiao $^{\orcidlink{0000-0002-6968-0919}}$, \IEEEmembership{Member, IEEE}, Xiaolu Kang $^{\orcidlink{0009-0009-2781-0891}}$, Qiguang Miao $^{\orcidlink{0000-0002-2872-388X}}$, \IEEEmembership{Senior Member, IEEE}, and Kun Xie $^{\orcidlink{0000-0003-4014-9646}}$
\thanks{This work has been submitted to the IEEE for possible publication. Copyright may be transferred without notice, after which this version may no longer be accessible.}
\thanks{The work was jointly supported by the National Science and Technology Major Project under Grant No. 2022ZD0117103, the National Natural Science Foundations of China under Grant No. 62272364, the provincial Key Research and Development Program of Shaanxi under Grant No. 2024GH-ZDXM-47, and the Research Project on Higher Education Teaching Reform of Shaanxi Province under Grant No. 23JG003. \textit{(Corresponding authors: Zhuoqi Ma and Qiguang Miao)}}
\thanks{Kang Liu, Zhuoqi Ma, Xiaolu Kang, Qiguang Miao, and Kun Xie are with the School of Computer Science and Technology, Xidian University, Xi'an, Shaanxi 710071, China, also with the Xi'an Key Laboratory of Big Data and Intelligent Vision, Xi'an, Shaanxi 710071, China, and also with the Key Laboratory of Collaborative Intelligence Systems, Ministry of Education, Xidian University, Xi'an 710071, China (e-mail: kangliu422@gmail.com; zhuoqima@xidian.edu.cn; lyisregret555@gmail.com;  qgmiao@xidian.edu.cn; xiekun@xidian.edu.cn). }
\thanks{Mengmeng Liu is with the School of Artificial Intelligence, Xidian University, Xi'an 710071, China (e-mail: mmengliu306@163.com). }
\thanks{Zhicheng Jiao is with the Department of Diagnostic Imaging, Brown University, Providence, RI 02903-4923, USA (e-mail: zhicheng\_jiao@brown.edu). }
}

\markboth{Journal of \LaTeX\ Class Files,~Vol.~14, No.~8, August~2021}%
{Shell \MakeLowercase{\textit{et al.}}: A Sample Article Using IEEEtran.cls for IEEE Journals}


\maketitle

\begin{abstract}
A radiology report comprises presentation-style vocabulary, which ensures clarity and organization, and factual vocabulary, which provides accurate and objective descriptions based on observable findings. While manually writing these reports is time-consuming and labor-intensive, automatic report generation offers a promising alternative. A critical step in this process is to align radiographs with their corresponding reports. However, existing methods often rely on complete reports for alignment, overlooking the impact of presentation-style vocabulary. To address this issue, we propose FSE, a two-stage Factual Serialization Enhancement method. In Stage 1, we introduce factuality-guided contrastive learning for visual representation by maximizing the semantic correspondence between radiographs and corresponding factual descriptions. In Stage 2, we present evidence-driven report generation that enhances diagnostic accuracy by integrating insights from similar historical cases structured as factual serialization. Experiments on MIMIC-CXR and IU X-ray datasets across specific and general scenarios demonstrate that FSE outperforms state-of-the-art approaches in both natural language generation and clinical efficacy metrics. Ablation studies further emphasize the positive effects of factual serialization in Stage 1 and Stage 2. The code is available at https://github.com/mk-runner/FSE.
\end{abstract}


\section{Introduction}
\IEEEPARstart{C}{hest} X-ray report generation aims to meticulously describe the morphological and positional features of essential organs and tissues in medical images using natural language. These reports play a critical role in assisting physicians with disease screening, diagnosis, and clinical decision-making. Nevertheless, composing these reports typically requires manual effort grounded on medical knowledge and clinical practical experiences, making the process both time-consuming and highly specialized \cite{lu2023effectively}. 

Automatic chest X-ray report generation (CXRG) \cite{sei,TriNet-2023-TMM,bannur2024maira2groundedradiologyreport,2024-mm-med-llm} offers a promising alternative by generating high-quality draft reports for radiologists, thereby enhancing diagnostic efficiency. To improve the clinical accuracy of generated reports, current predominant approaches in CXRG encompass cross-modal alignment methods \cite{chen-etal-2020-generating,Qin_reinfor_acl_2022,wang-cross-modal,Panoramic-2024-TMM}, knowledge-enhanced strategies \cite{yang-gsket,huang-kiut,wang2023metransformer,yang-m2kt}, region-guided techniques \cite{recap,tanida-rgrg,2022-miccai-region}, and large language model-based approaches \cite{yan2023style,aaai-liu2024bootstrapping-llm,2024-mm-med-llm,lu2023effectively}. However, CXRG still faces several challenges: \textbf{1) Neglecting the impact of presentation-style vocabulary on cross-modal alignment.} Radiology reports comprise presentation-style vocabulary, which ensures clarity and organization, and factual vocabulary, which provides accurate and objective descriptions based on observable findings. Although the latter is evidently more important than the former in clinical practice, existing methods \cite{wang-mgca,yang-m2kt,align-2024-TMM} still rely on complete reports for cross-modal alignment. \textbf{2) Difficulty in extending existing knowledge retrieval methods to other datasets.} These methods \cite{yang-gsket,li-towards} depend on disease labels, restricting their broader applicability. \textbf{3) Challenges in evaluating completeness between generated and reference reports during testing.} Concise medical reports are effective in specific scenarios, like emergency diagnoses, leading many existing methods \cite{chen-etal-2020-generating,sei,2024-eccv-hergen} to truncate reference reports to a fixed length.  However, this evaluation method may omit critical findings, making it unsuitable for general scenarios requiring comprehensive assessment.

To address these challenges, we introduce a novel two-stage method called \textbf{F}actual \textbf{S}erialization \textbf{E}nhancement (FSE) for chest X-ray report generation. In the first stage, we propose factuality-guided contrastive learning for visual representation by maximizing the semantic correspondence between radiographs and factual serialization derived from corresponding reports. Specifically, we develop a structural entities approach to generate factual serialization (i.e., concise sentences consisting solely of factual vocabulary from reports) by excluding presentation-style vocabulary. We then align radiographs with factual serialization through instance-level and token-level cross-modal semantic correspondences. In the second stage, we present evidence-driven report generation that enhances diagnostic accuracy by incorporating insights from similar historical cases structured as factual serialization. First, we retrieve similar historical cases for each radiograph based on semantic similarities between aligned unimodal representations, independent of disease labels. Next, we employ a cross-modal fusion network to integrate these insights, guiding the text generator to produce clinically accurate reports. Experiments results on MIMIC-CXR and IU X-ray datasets demonstrate that FSE outperforms recent state-of-the-art methods in both specific and general scenarios. Additionally, factual serialization plays critical roles in factuality-guided contrastive learning and evidence-driven report generation, leading us to term our method as factual serialization enhancement. Our key contributions are summarized as follows: 
\begin{itemize}
    \item We propose factuality-guided contrastive learning, effectively aligning radiographs with corresponding factual descriptions.
    \item We present evidence-driven report generation, enhancing diagnostic accuracy by integrating insights from similar historical cases.
    \item We conduct comprehensive evaluations of FSE in specific and general scenarios, demonstrating its superiority over existing methods.
\end{itemize}

\begin{figure*}
    \centering
    \includegraphics[width=1\linewidth]{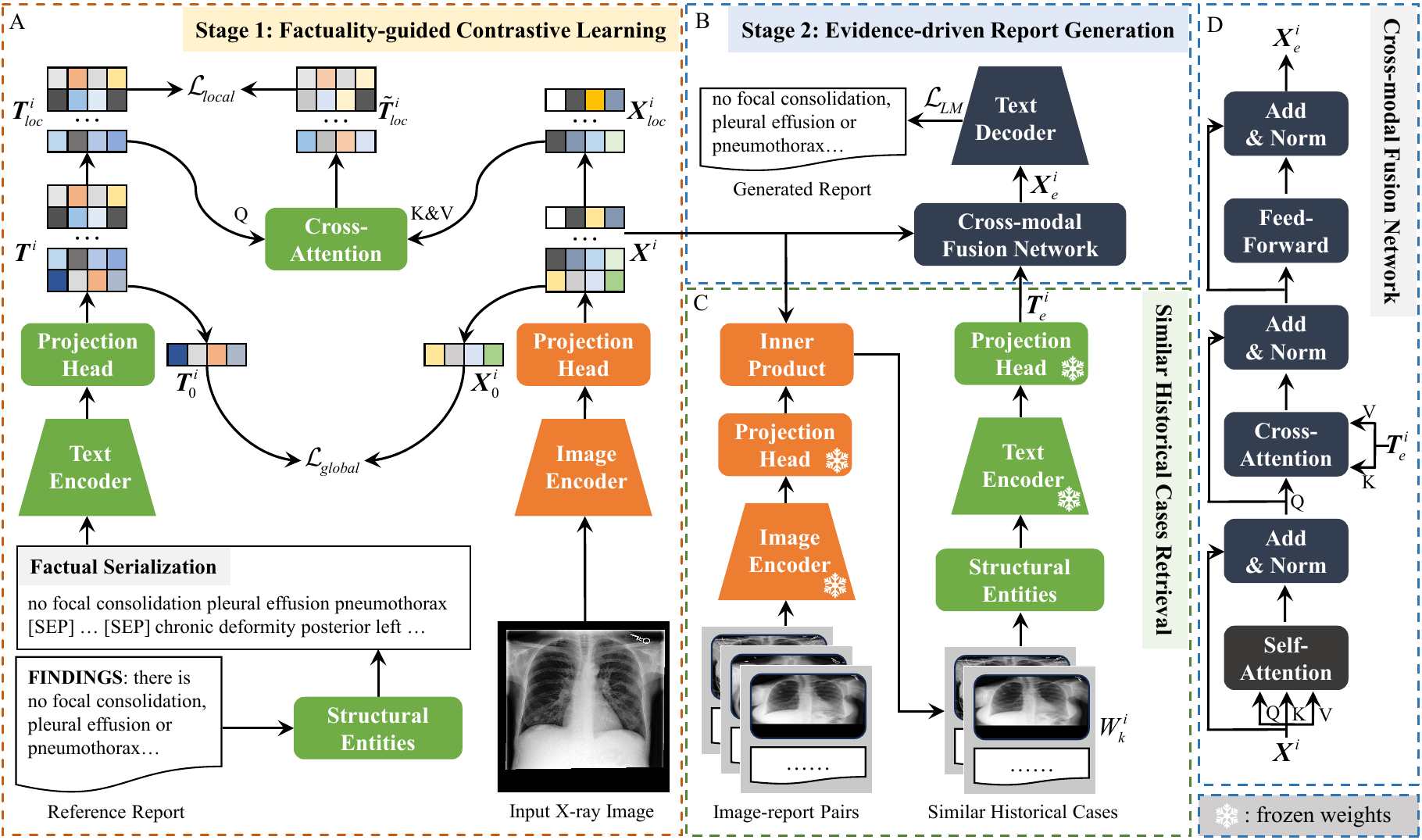}
    \caption{Overview of our proposed FSE for chest X-ray report generation. FSE involves factuality-guided contrastive learning (A) for visual representation, followed by gradient-free retrieval of similar historical cases (B), and concludes with evidence-driven report generation. The inference phase exclusively employs evidence-driven report generation.}
    \label{fig:1}
\end{figure*}

\section{Related Work}
We briefly review related work about cross-modal alignment with image-report pairs and retrieval-based chest X-ray report generation.

\textbf{Cross-modal alignment with image-report pairs.} This technique is crucial in medical image analysis, as it enhances unimodal representations by maximizing consistency between matching image-report pairs while pushing apart unrelated pairs. Existing methods \cite{medclip_wang_2022,zhang-kad,chen-ptunifier,wang-mgca} typically rely on complete reports for cross-modal alignment to learn generalized medical visual representations. For instance, PTUnifier \cite{chen-ptunifier} performs instance-wise and token-wise alignment to match radiographs with corresponding reports. MGCA \cite{wang-mgca} utilizes natural semantic correspondences between image-report pairs to perform multi-grained alignments across the instance, pathological region, and disease levels. MedCLIP \cite{medclip_wang_2022} applies a semantic matching loss to decouple images and reports, facilitating more effective multimodal contrastive learning. However, these approaches often overlook the impact of presentation-style vocabulary, which is less relevant to clinical practice compared to factual vocabulary. Therefore, we propose factuality-guided contrastive learning, focusing on aligning radiographs with factual serialization to extract clinically valuable visual features.

\textbf{Retrieval-based chest X-ray report generation.} This approach \cite{Liu_2021_CVPR_retrieve, pmlr-v158-endo21a-retrieve-ML4H,pmlr-v227-jeong24a-MIDL-retrieve} selects the most similar reports or sentences from a large corpus, such as the training set, to serve as generate reports. For example, CXR-RePaiR \cite{pmlr-v158-endo21a-retrieve-ML4H} employs a vision-language pre-trained model, such as CLIP \cite{radford-learning}), to compute the similarities between the image and text, choosing the most similar report as the output. X-REM \cite{pmlr-v227-jeong24a-MIDL-retrieve} first leverages the ALBEF model \cite{li-albef} to retrieve the top \textit{i} reports that are most similar to the input radiographs and then refines these results by incorporating image-report matching and natural language inference scores, thereby producing a retrieval-based report. While retrieved reports offer prior knowledge, they may not perfectly match the current patient, impacting report quality. Therefore, we propose evidence-driven report generation that employs the cross-attention mechanism \cite{NIPS2017_transformer_attention} to integrate similar historical cases with the current patient's radiograph, incorporating relevant prior knowledge and patient radiological data into the report.

\section{Methodology}
\label{sec: methods}
Fig. \ref{fig:1} illustrates an overview of our proposed FSE. In Stage 1, we introduce factuality-guided contrastive learning, which focuses on capturing clinically relevant visual features that are then used to retrieve similar historical cases. In Stage 2, we propose evidence-driven report generation, which integrates these retrieved cases with the patient’s radiological data. The following subsections provide a detailed explanation of the proposed factuality-guided contrastive learning and evidence-driven report generation.

\subsection{Problem Formulation}
Let \({{\cal D}_{train}} = \left\{ {\left. {\left( {{x^i},{y^i}} \right)} \right|{x^i} \in {\cal X},{y^i} \in {\cal Y}}, {1 \le i \le n} \right\}\) be the training set with \textit{n} samples, where \({\cal X}\) and \({\cal Y}\) are the sets of radiographs and reports, respectively. The test set \({{\cal D}_{test}} = \left\{ {\left. {{x^j}} \right|{x^j} \in {\cal X},1 \le j \le m} \right\}\) consists of \textit{m} samples. This work aims to train a model \({F_\theta }:\left( {{\cal X},{\cal W}} \right) \to {\cal Y}\) with parameters \(\theta\) using \({\cal D}_{train}\) and then generate a report for each radiograph in \({{\cal D}_{test}}\). Here, \({\cal W} = \left\{ {W_k^1, \cdots ,W_k^n, \cdots ,W_k^{n + m}} \right\}\)\footnote{To prevent data leakage, each sample's similar historical cases exclude the sample itself.} denotes the set of similar historical cases per sample, where \(W_k^l = \left\{ {\left. {\left( {{x^o},{y^o}} \right)} \right|\left( {{x^o},{y^o}} \right) \in {{\cal N}_k} \left( {{x^l},{y^l}} \right) \subseteq {{\cal D}}_{train}} \right\}\) refers to similar historical cases of the \(l^{th}\) sample, and \({{\cal N}_k}\left ( {{x^l},{y^l}} \right ) \) denotes the top \textit{k} most similar samples in training set to \(\left({x^l,y^l} \right)\).

\subsection{Factuality-guided Contrastive Learning} \label{sec: cross-modal alignment}
In this subsection, we introduce the structural entities approach for extracting factual serialization from reports, describe the extraction of unimodal features via image and text encoders, and present factuality-guided contrastive learning.

\begin{figure}
    \centering
    \includegraphics[width=1\linewidth]{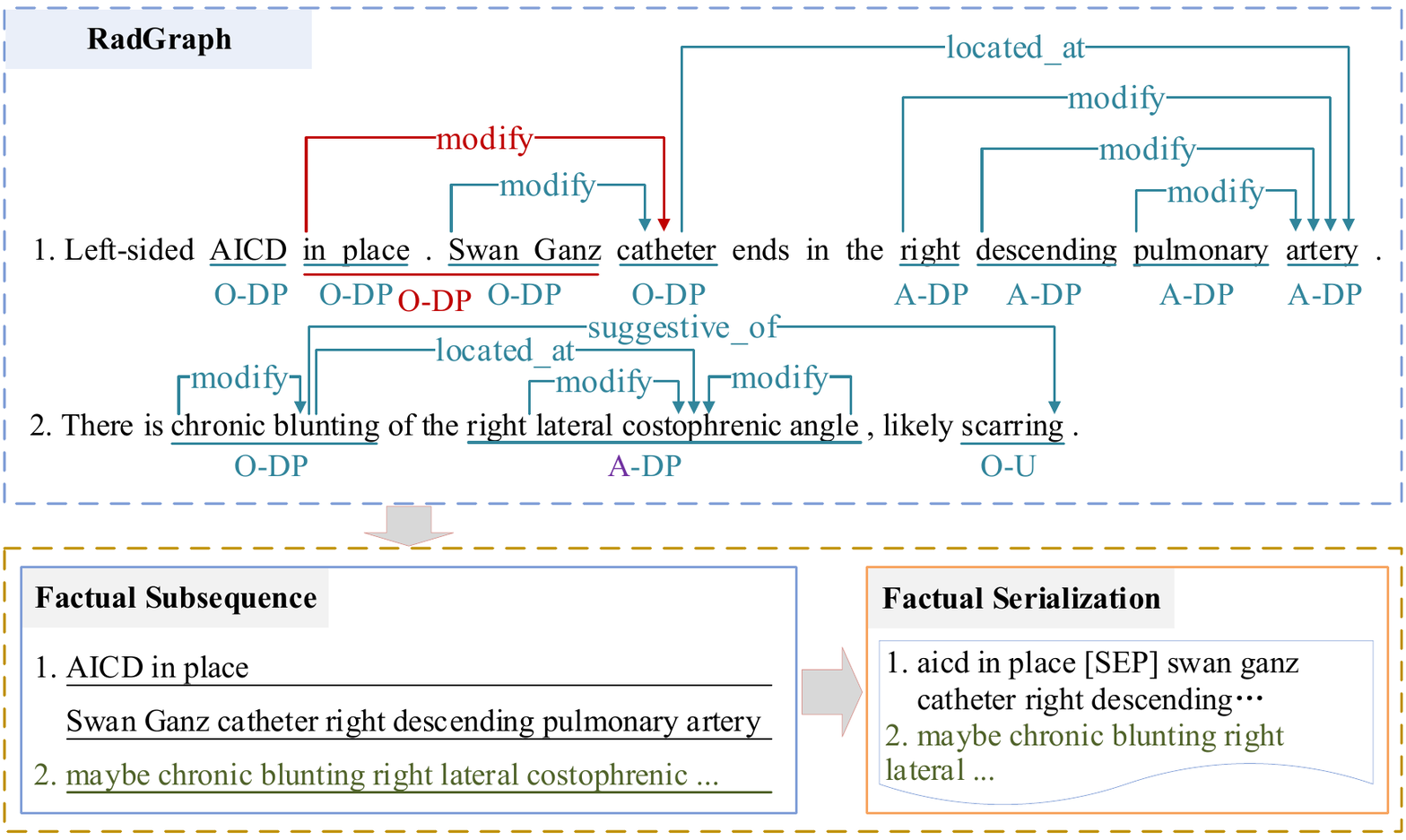}
    \caption{Two examples of factual serialization generated by our structural entities approach. The upper panel shows entities and their relationships identified by RadGraph \cite{jain-radgraph}, while the lower panel shows factual serialization from reports. “O-DP”, "A-DP", and "O-U" indicate entity types. “modify”, “located\_at”, and “suggestive\_of” denote relationships between them.}
    \label{fig:2}
\end{figure}

\textbf{Structural entities approach for extracting factual serialization from reports.} Clinical decision-making relies heavily on accurate and fact-based descriptions, but existing methods \cite{medclip_wang_2022,song-etal-2022-cross,wang-mgca} directly utilize complete reports for cross-modal alignment. While presentation-style vocabulary enhances the readability and structure of reports, it can hinder effective image-report alignment. To overcome this, we propose the Structural Entities (SE) approach, which extracts factual serialization consisting solely of factual vocabulary from reports, enabling the model to focus on aligning radiographs with the corresponding factual serialization (see Fig. \ref{fig:2})

Our SE approach builds upon the outputs of RadGraph \cite{jain-radgraph}, a tool that extracts entities and relations from full-text radiology reports, such as CT, PET, and X-ray. Although RadGraph effectively identifies a broad range of entities and their interrelations, its output may include noise, such as entities spanning two sentences (e.g., “\textit{in place . Swan Ganz}”) or clinically insignificant single-word entities (e.g., “\textit{These}”). To mitigate this issue, we first remove such entities and resolve overlapping entities by retaining only the longest and most informative instance (e.g., selecting  “\textit{1.9 \(\times \) 1.0 cm}” over “\textit{1.0 cm}”). Subsequently, the remaining entities are organized in their original order within the report to preserve contextual integrity and segmented into factual subsequences based on sentence-ending punctuation. For subsequences containing entities labeled as “Definitely Absent” (O-DA) or “Uncertain” (O-U), indicators such as “no” or “maybe” are prefixed to clarify the factual context (see Fig. \ref{fig:2}). Finally, subsequences are connected using the [SEP] token to generate factual serialization for the \textit{i}\textsuperscript{th} sample:
\begin{equation}
{S^i} = \left\{ {s_1^i,[{\rm{SEP}}],s_2^i,[{\rm{SEP}}], \cdots ,[{\rm{SEP}}],s_{N_s^i}^i} \right\},
\end{equation}
where \(N_s^i\) denotes the number of factual subsequences. Although both our SE method and \cite{yan2023style} derive factual serialization from reports using RadGraph \cite{jain-radgraph} outputs, two key differences exist: First, we refine RadGraph outputs by removing noisy entities, whereas \cite{yan2023style} directly utilizes the raw outputs. Second, our approach preserves the original sequence of entities as they appear in the report, in contrast to \cite{yan2023style}, which relies on the graph structure. The latter may produce multiple disconnected subgraphs for entities within the same sentence, potentially obscuring their interrelationships and distorting the original semantic content.

\textbf{Extraction of unimodal features via image and text encoders.} In line with established practices \cite{chen-etal-2020-generating,chen-etal-2021-cross-modal,yang-gsket}, we use ResNet101 \cite{he-resnet}, pre-trained on ImageNet, as the image encoder. Visual features \({\boldsymbol{X}}\in {\mathbb{R}^{B \times 50 \times 2048}}\) are extracted from the final convolutional layer: 
\begin{equation}
{\boldsymbol{X}}=\left [ {{\boldsymbol{X}_0},{\boldsymbol{X}_1}, \cdots ,{\boldsymbol{X}_{49}}} \right ]  = G_{{\theta _I}}\left( {{\cal X}} \right),
\end{equation}
where \(\boldsymbol{X}_{0}\) denotes global visual features obtained via global averaging pooling, and \(G_{{\theta _I}}\left(  \cdot  \right)\) is the image encoder with parameters \({\theta _I}\). \textit{B} refers to the batch size in the mini-batch.

Before extracting textual features for factual serialization, we train a word-level tokenizer based on the training corpus to encode factual serialization \(S = \left\{ {{S^1}, {S^2}, \cdots ,{S^B}} \right\}\). Textual features \({\boldsymbol{T}}\in {\mathbb{R}^{B \times \left ( N_R+1 \right ) \times 768}}\) are then extracted using a six-layer pre-trained SciBERT \cite{Beltagy2019} model:
\begin{equation}
{\boldsymbol{T}}=\left [ {{{\boldsymbol{T}}_0},{{\boldsymbol{T}}_1}, \cdots ,{{\boldsymbol{T}}_{{N_R}}}} \right ] = H_{{\theta _R}}\left( S \right),
\end{equation}
where \({N_R}\) indicates the number of tokens, and \({\boldsymbol{T}_0}\) represents global textual features extracted by the text encoder \(H_{{\theta _R}}\left(  \cdot  \right)\) with parameters \({\theta_R}\). Notably, \({\boldsymbol{T}}\) is used only in the first stage.

For the convenience of aligning visual and textual features in a common space, we use projection heads to map these features to a unified dimensionality, labeled as \textit{d}. For simplicity, we denote the features after projection as \({\boldsymbol{X}}\in {\mathbb{R}^{B \times 50 \times d}}\) for visual features and \({\boldsymbol{T}}\in {\mathbb{R}^{B \times \left ( N_R+1 \right )  \times d}}\) for textual features.

\textbf{Factuality-guided contrastive learning.} To enhance the model's understanding of medical images, we introduce factuality-guided contrastive learning that leverages the natural semantic correspondence between image-report pairs at both instance and toke levels for improved visual representation. We detail the factuality-guided instance alignment and factuality-guided token alignment as follows. 

To ensure consistent representations between radiographs and corresponding factual serialization, we define the report-to-image factuality-guided instance alignment loss as follows:
\begin{align}
{{\cal L}}_{global}^{I \leftarrow R} =  - \frac{1}{B}\sum\limits_{i = 1}^B {\log \frac{{\exp \left( \text{sim}\left ({\boldsymbol{X}}_{0}^i, {\boldsymbol{T}}_{0}^i \right )  \mathord{\left/ \right.} {{\tau_1}} \right)}}{{\sum\nolimits_{j = 1}^B {\exp \left( \text{sim}\left ({\boldsymbol{X}}_{0}^i, {\boldsymbol{T}}_{0}^j \right ) \mathord{\left/ \right.} {{\tau}_1} \right)} }}},
\end{align}
where \(\tau_1\) means the temperature parameter. \({\boldsymbol{X}_{0}^i}\) and \({\boldsymbol{T}_{0}^i}\) denote global features of medical images and factual serialization for the \(i^{th}\) sample, respectively. The function \(\text{sim} \left ( \boldsymbol{a} , \boldsymbol {b} \right ) = (\boldsymbol {a} \cdot \boldsymbol {b}^T) /\left ( \left \| \boldsymbol {a} \right \| \cdot \left \| \boldsymbol {b} \right \|  \right )   \) represents cosine similarity between \( \boldsymbol{a}\) and \( \boldsymbol{b}\). The image-to-report factuality-guided instance alignment loss \({{\cal L}}_{global}^{R \leftarrow I}\) is defined similarly and serves as the symmetric counterpart to \({{\cal L}}_{global}^{I \leftarrow R}\). Thus, the overall loss \({\cal L}_{global}\) is given by \({{\cal L}}_{global}=0.5\times ( {{\cal L}}_{global}^{I \leftarrow R}+{{\cal L}}_{global}^{R \leftarrow I} ) \).

Recognizing the significance of fine-grained features in medical image analysis, we introduce the factuality-guided token alignment. We first derive the local cross-modal textual features \({{{\tilde {\boldsymbol{T}}}}_{loc}} \in {{\mathbb{R} }^{B \times {N_R} \times d}}\) through the cross-attention mechanism \cite{NIPS2017_transformer_attention}, denoted as \({\mathop{\rm ATTN}\nolimits}\left ( Q,K,V \right ) \). This process involves treating local textual features \({\boldsymbol{T}_{loc}}=\left [ {{{\boldsymbol{T}}_1}, \cdots ,{{\boldsymbol{T}}_{{N_R}}}} \right ]\) as queries, and local visual features \({\boldsymbol{X}_{loc}}=\left [ {{{\boldsymbol{X}}_1}, \cdots ,{{\boldsymbol{X}}_{{49}}}} \right ]\) as keys and values to output the local cross-modal textual features, formulated as:
\begin{equation}
    \begin{aligned}
          {{{\tilde {\boldsymbol{T}}}}_{loc}} &= {\mathop{\rm ATTN}\nolimits} \left( {{{\boldsymbol{T}}_{loc}},{{\boldsymbol{X}}_{loc}},{{\boldsymbol{X}}_{loc}}} \right) \\
           &= {\mathop{\rm softmax}\nolimits} \left( {{{{{\boldsymbol{T}}_{loc}} \cdot {{\boldsymbol{X}}_{loc}^T}} / {\sqrt d }}} \right) \cdot {{\boldsymbol{X}}_{loc}}
    \end{aligned}
\end{equation}
Due to the potential sharing of similar semantic information in reports from different samples, determining negative pairs becomes challenging. Therefore, within a sample, we designate tokens at the same position as positive pairs and those at different positions as negative pairs. The factuality-guided token alignment loss is defined as:
\begin{equation}
    \begin{aligned}
        {{\cal L}}_{local}^i =  - \frac{1}{{2{N_R}}}\sum\limits_{v = 1}^{{N_R}} {\log \frac{{\exp \left( \text{sim}\left ({\boldsymbol{T}}_{loc}^{i,v}, {\tilde{\boldsymbol{T}}}_{loc}^{i,v} \right )  \mathord{\left/ \right.} {{\tau_2}} \right)  }}{{\sum\nolimits_{u = 1}^{{N_R}} {\exp \left( \text{sim}\left ({\boldsymbol{T}}_{loc}^{i,v}, {\tilde{\boldsymbol{T}}}_{loc}^{i,u} \right )  \mathord{\left/ \right.} {{\tau_2}} \right)} }}} \\
         - \frac{1}{{2{N_R}}}\sum\limits_{v = 1}^{{N_R}}{\log \frac{{\exp \left( \text{sim}\left ({\tilde{\boldsymbol{T}}}_{loc}^{i,v}, {\boldsymbol{T}}_{loc}^{i,v}\right )  \mathord{\left/ \right.} {{\tau_2}} \right)}}{{\sum\nolimits_{u = 1}^{{N_R}} {\exp \left( \text{sim}\left ({\tilde{\boldsymbol{T}}}_{loc}^{i,v}, {\boldsymbol{T}}_{loc}^{i,u}\right )  \mathord{\left/ \right.} {{\tau_2}} \right)} }}},
    \end{aligned}
\end{equation}
where \({N_R}\) and \(\tau_2\) denote the number of tokens for textual features and the temperature parameter, respectively. \({\boldsymbol{T}}_{loc}^{i,v} \in {{\mathbb{R} }^{d}}\) represents the local textual feature of the \(v^{th}\) token in the \(i^{th}\) sample. \({{\cal L}}_{local} = {1 \mathord{\left/ {\vphantom {1 B}} \right.} B}\sum\nolimits_{i = 1}^B {{{\cal L}}_{local}^i}\). 

Significantly, while both our method and MGCA \cite{wang-mgca} leverage instance-level and token-level semantic correspondences between image-report pairs for visual representation, the key distinction is our use of factual serialization from reports for cross-modal alignment, rather than using the complete report. This approach enables the model to focus on aligning radiographs with corresponding factual descriptions, thereby reducing the influence of presentation-style vocabulary on the alignment. Furthermore, our method supports the retrieval of cases with similar factual descriptions based on the aligned visual features, without requiring disease labels, as detailed in Section \ref{sec: report_generation_module}. To summarize, the training objective for factuality-guided contrastive learning is formulated as: 
\begin{align}
    {{{\cal L}}_{pretrain}} = {{\cal L}}_{global}  + {{\cal L}}_{local}.
\end{align}

\begin{table*}
\centering
\caption{Comparisons with SOTA methods on MIMIC-CXR. \({\spadesuit}\) denotes results from the original work, excluding BS and CX5, which were not reported. Other results are reproduced using the official code. The best values for each \({\boldsymbol{M_{gt}}}\) are in \textbf{bold}. }
\label{table: main results}
\begin{tabular}{llccccccccccc} 
\toprule
\multirow{2}{*}{\textbf{Method}} & \multirow{2}{*}{\textbf{Venue}} & \multirow{2}{*}{${\boldsymbol{M_{gt}}}$} & \multicolumn{7}{c}{\textbf{NLG}$\uparrow$} & \multicolumn{3}{c}{\textbf{CE}$\uparrow$} \\ 
\cmidrule(lr){4-10}\cmidrule(lr){11-13}
 &  &  & \textbf{BS} & \textbf{BL-1} & \textbf{BL-2} & \textbf{BL-3} & \textbf{BL-4} & \textbf{MTR} & \textbf{R-L} & \textbf{RG} & \textbf{CX5} & \textbf{CX14} \\ 
\midrule
\multicolumn{13}{c}{Retrieval-based approaches} \\ 
\midrule
X-REM & MIDL’23 & -$^{\spadesuit}$ & - & - & 0.186 & - & - & - & - & 0.181 & - & 0.381 \\ 
\midrule
\multicolumn{13}{c}{Generative-based  approaches} \\ 
\midrule
\multirow{2}{*}{R2Gen} & \multirow{2}{*}{EMNLP’20} & 100$^{\spadesuit}$ & 0.841 & {0.353} & 0.218 & 0.145 & 0.103 & 0.142 & 0.277 & 0.207 & 0.340 & {0.340} \\
 &  & \textit{Cpl.} & 0.841 & 0.350 & 0.209 & 0.137 & 0.097 & 0.135 & 0.266 & 0.211 & 0.339 & 0.338 \\ 
\hdashline[1pt/1pt]
\multirow{2}{*}{R2GenCMN} & \multirow{2}{*}{ACL’21} & 100$^{\spadesuit}$ & 0.843 & {0.353} & 0.218 & 0.148 & 0.106 & 0.142 & 0.278 & {0.220} & {0.461} & 0.278 \\
 &  & \textit{Cpl.} & {0.844} & 0.328 & 0.198 & 0.130 & 0.090 & 0.133 & 0.268 & 0.223 & 0.464 & 0.393 \\ 
\hdashline[1pt/1pt]
\multirow{2}{*}{CvT2DistillGPT2} & \multirow{2}{*}{ARTMED’23} & 60$^{\spadesuit}$ & 0.839 & \textbf{0.393} & \textbf{0.248} & \textbf{0.171} & \textbf{0.127} & 0.155 & \textbf{0.286} & {0.223} & {0.463} & {0.391} \\
 &  & \textit{Cpl.} & 0.842 & 0.323 & 0.204 & 0.140 & 0.102 & {0.138} & 0.277 & {0.237} & {0.483} & {0.434} \\ 
\hdashline[1pt/1pt]
\multirow{2}{*}{M2KT} & \multirow{2}{*}{MedIA’23} & 80$^{\spadesuit}$ & 0.831 & 0.386 & 0.237 & 0.157 & 0.111 & 0.137 & 0.274 & {0.204} & {0.477} & 0.352 \\
 &  & \textit{Cpl.} & 0.833 & {0.351} & 0.204 & 0.128 & 0.085 & 0.133 & 0.244 & 0.210 & {0.483} & 0.413 \\ 
\hdashline[1pt/1pt]
GSKET & MedIA’22 & 80$^{\spadesuit}$ & - & 0.363 & 0.228 & 0.156 & 0.115 & - & \textbf{0.284} & - & - & {0.371} \\
CMCA & COLING’22 & -$^{\spadesuit}$ & - & 0.360 & 0.227 & 0.156 & 0.117 & 0.148 & 0.287 & - & - & 0.356 \\
DCL & CVPR’23 & 90$^{\spadesuit}$ & - & - & - & - & 0.109 & 0.150 & \textbf{0.284} & - & - & {0.373} \\
METransformer & CVPR’23 & -$^{\spadesuit}$ & - & 0.386 & 0.250 & 0.169 & 0.124 & 0.152 & 0.291 & - & - & 0.311 \\
OpenLLaMA-7B & NIPSW’23 & -$^{\spadesuit}$ & - & - & - & - & 0.069 & - & 0.235 & - & 0.422 & 0.320 \\
SA & EMNLP’23 & -$^{\spadesuit}$ & - & - & 0.162 & - & - & - & - & 0.228 & - & 0.394 \\
MMTN & AAAI’23 & -$^{\spadesuit}$ & - & 0.379 & 0.238 & 0.159 & 0.116 & - & 0.283 & - & - & - \\
CAMANet & JBHI’24 & -$^{\spadesuit}$ & - & 0.374 & 0.230 & 0.155 & 0.112 & 0.145 & 0.279 & - & - & 0.387 \\ 
RAMT & TMM’24 & -$^{\spadesuit}$ & - & 0.362 & 0.229 & 0.157 & 0.113 & 0.153 & 0.284 & - & - & 0.335 \\ 
Pretrain+FMVP & TMM’24 & -$^{\spadesuit}$ & - & 0.389 & 0.236 & 0.156 & 0.108 & 0.150 & 0.284 & - & - & 0.336 \\ 
\midrule
\multirow{5}{*}{\textbf{FSE-1 (Ours)}} & \multirow{5}{*}{-} & 60 & \textbf{0.841} & {0.381} & {0.232} & {0.156} & {0.113} & \textbf{0.158} & {0.279} & \textbf{0.225} & \textbf{0.499} & \textbf{0.453} \\
 &  & 80 & \textbf{0.844} & \textbf{0.397} & \textbf{0.242} & \textbf{0.163} & \textbf{0.117} & \textbf{0.155} & {0.281} & \textbf{0.238} & \textbf{0.529} & \textbf{0.470} \\
 &  & 90 & \textbf{0.845} & \textbf{0.389} & \textbf{0.238} & \textbf{0.160} & \textbf{0.115} & \textbf{0.153} & {0.281} & \textbf{0.240} & \textbf{0.536} & \textbf{0.472} \\
 &  & 100 & \textbf{0.845} & \textbf{0.384} & \textbf{0.235} & \textbf{0.158} & \textbf{0.113} & \textbf{0.152} & \textbf{0.281} & \textbf{0.241} & \textbf{0.538} & \textbf{0.473} \\
 &  & \textit{Cpl.} & \textbf{0.845} & \textbf{0.373} & \textbf{0.228} & \textbf{0.152} & \textbf{0.109} & \textbf{0.148} & \textbf{0.279} & \textbf{0.241} & \textbf{0.542} & \textbf{0.474} \\ 
\bottomrule
\end{tabular}
\end{table*}

\subsection{Evidence-driven Report Generation} \label{sec: report_generation_module}
This subsection introduces the retrieval of similar historical cases, followed by evidence-driven report generation.

\textbf{Retrieval of similar historical cases.} Radiologists may reference similar cases to accurately identify imaging features and improve diagnostic precision \cite{wang2024-survey-rrg}. Consequently, retrieving high-quality similar historical cases is crucial for effective report generation. These similar cases are image-report pairs from the training set that closely match the current radiograph. Existing retrieval methods typically rely on image-text similarity \cite{pmlr-v158-endo21a-retrieve-ML4H}, image-report matching scores \cite{pmlr-v227-jeong24a-MIDL-retrieve}, or shared disease labels \cite{yang-gsket,li-towards}. However, these approaches may encounter modality gaps or require additional disease labels for effective retrieval. To overcome these challenges, we retrieve similar historical cases by focusing on the similarity of aligned visual features, rather than relying on traditional image-text similarity, image-report matching scores, or disease labels. This method circumvents modality gaps and the need for extra disease labels. To improve retrieval efficiency, we employ the Faiss \cite{johnson2019billion} tool to compute visual feature similarities and conduct the gradient-free retrieval process only once.

\textbf{Evidence-driven report generation.} To emulate the practice of radiologists referencing similar historical cases during diagnosis, we propose evidence-driven report generation. This approach enhances diagnostic accuracy by integrating insights from similar historical cases, structured as factual serialization. First, similar historical cases for the \(i^{th}\) sample are sequentially processed through the frozen text encoder and projection head to generate evidence features \({\boldsymbol{T}^i_e} \in {\mathbb{R}^{ \left ( N_{ke}^i+1 \right ) \times d}}\):
\begin{equation}
{\boldsymbol{T}^i_e} = H_{{\theta _R}}\left( \text{SE}\left ( W_k^i \right )   \right),
\end{equation}
where \(\text{SE}\left ( \cdot  \right ) \) refers to our structural entities approach described in Section \ref{sec: cross-modal alignment}. \({N_{ke}^i}= {\textstyle \sum_{u=1}^{k}} N_u^i\) denotes the total number of tokens in the local evidence features, with \(N_u^i\) indicating the number of tokens in the \(u^{th}\) similar cases for the \(i^{th}\) sample. When each sample has multiple similar historical cases (\(k > 1\)), their textual features are concatenated to form evidence features. To effectively integrate these features, we devise the cross-modal fusion network (CMF) based on a Transformer Decoder layer \cite{chen2022align, chen-ptunifier}, as shown in Fig. \ref{fig:1}(D). The network comprises a self-attention sub-layer, cross-attention sub-layer, and feed-forward sub-layer, formulated as:
\begin{equation}
    \begin{aligned}
          {{\boldsymbol{X}}_{s}^{i}} &= \text{LN}\left ({\boldsymbol{X}}^{i} + {\mathop{\rm ATTN}\nolimits} \left( {{{\boldsymbol{W}}_{qs}{\boldsymbol{X}}^{i}},{{\boldsymbol{W}}_{ks}{\boldsymbol{X}}^{i}},{{\boldsymbol{W}}_{vs}{\boldsymbol{X}}^{i}}} \right) \right ) \\  
          {{\boldsymbol{X}}_{c}^{i}} &= \text{LN}\left ({\boldsymbol{X}}_{s}^{i} + {\mathop{\rm ATTN}\nolimits} \left( {{{\boldsymbol{W}}_{qc}{\boldsymbol{X}}_{s}^{i}},{{\boldsymbol{W}}_{kc}{\boldsymbol{T}}_{e}^{i}},{{\boldsymbol{W}}_{vc}{\boldsymbol{T}}_{e}^{i}}} \right) \right ) \\
          {{\boldsymbol{X}}_{e}^{i}} &= \text{LN}\left ({\boldsymbol{X}}_{c}^{i} + {\mathop{\rm FFN}\nolimits} \left( {\boldsymbol{X}}_{c}^{i} \right) \right ), \\  
    \end{aligned}
\end{equation}
where \({\boldsymbol{W}}_{qs}\), \({\boldsymbol{W}}_{ks}\), \({\boldsymbol{W}}_{vs}\), \({\boldsymbol{W}}_{qc}\), \({\boldsymbol{W}}_{kc}\), and \({\boldsymbol{W}}_{vc}\) are learnable parameters. \(\text{LN}\left ( \cdot  \right ) \) and \(\text{FFN}\left ( \cdot  \right ) \) represent the layer normalization and feed-forward sub-layer, respectively. Finally, the text decoder generates reports autoregressively, conditioned on the multi-modal representations \({{\boldsymbol{X}}_{e}^{i}}\). The evidence-driven report generation module with parameters \({\theta _D}\) is optimized by minimizing the negative log-likelihood \( {P_{{\theta _D}}}\left( {\left. {y_t^i} \right|{y_{<t}^i, {{\boldsymbol{X}}_{e}^{i}}}} \right)\): 
\begin{equation}
    {{{\cal L}}_{LM}} =  - \frac{1}{B}\sum\limits_{i = 1}^B {\sum\limits_{t = 1}^M {\log {P_{{\theta _D}}}\left( {\left. {y_t^i} \right|{y_{<t}^i,{{\boldsymbol{X}}_{e}^{i}}}}\right)} } ,
\end{equation}
where \textit{M }represents the maximum length of tokens generated by the text decoder. \(y_{<t}\) denotes the previously predicted word sequence. Significantly, while our proposed FSE method shares certain similarities with retrieval-based approaches \cite{pmlr-v158-endo21a-retrieve-ML4H,pmlr-v227-jeong24a-MIDL-retrieve}, it differs in a key aspect: Unlike directly using similar historical cases as reports for given images, our method utilizes the cross-modal fusion network to extract relevant empirical features from these cases. The text decoder then generates reports based on these empirical and visual features.

\begin{table*}
\centering
\caption{Comparisons with SOTA methods on IU X-ray. \({\spadesuit}\) denotes results from the published literature, excluding BS, CX5, and CX14, which were not reported. The remaining results are reproduced using the official code and checkpoints. For each \({\boldsymbol{M_{gt}}}\), the best and second-best values are in \textbf{bold} and \underline{underlined}, respectively.}
\label{table: main-results-iu}
\begin{tabular}{llccccccccccc} 
\toprule
\multirow{2}{*}{\textbf{Method}} & \multirow{2}{*}{\textbf{Venue}} & \multirow{2}{*}{\({\boldsymbol{M_{gt}}}\)} & \multicolumn{7}{c}{\textbf{NLG}${\uparrow}$} & \multicolumn{3}{c}{\textbf{CE}$\uparrow$} \\ 
\cmidrule(lr){4-10}\cmidrule(lr){11-13}
 &  &  & \textbf{BS} & \textbf{BL-1} & \textbf{BL-2} & \textbf{BL-3} & \textbf{BL-4} & \textbf{MTR} & \textbf{R-L} & \textbf{RG} & \textbf{CX5} & \textbf{CX14} \\ 
\midrule
\multirow{2}{*}{R2Gen} & \multirow{2}{*}{EMNLP’20} & 60$^{\spadesuit}$ & 0.866 & 0.470 & 0.304 & 0.219 & 0.165 & 0.187 & 0.371 & 0.351 & 0.000 & 0.533 \\
 &  & \textit{Cpl.} & 0.864 & 0.465 & 0.302 & 0.216 & 0.162 & 0.201 & 0.364 & 0.349 & 0.000 & 0.531 \\ 
\hdashline[1pt/1pt]
\multirow{2}{*}{R2GenCMN} & \multirow{2}{*}{ACL’21} & 60$^{\spadesuit}$ & \underline{0.868} & 0.475 & 0.309 & 0.222 & 0.170 & 0.191 & 0.375 & \textbf{0.384} & 0.052 & 0.543 \\
 &  & \textit{Cpl.} & \textbf{0.867} & 0.471 & 0.306 & 0.219 & 0.166 & \underline{0.206} & 0.373 & \textbf{0.382} & 0.052 & 0.537 \\ 
\hdashline[1pt/1pt]
\multirow{2}{*}{CvT2DistillGPT2} & \multirow{2}{*}{ARTMED’23} & 60$^{\spadesuit}$ & \underline{0.868} & 0.473 & 0.304 & 0.224 & \underline{0.175} & 0.200 & 0.376 & 0.351 & 0.005 & 0.543 \\
 &  & \textit{Cpl.} & \underline{0.866} & 0.334 & 0.203 & 0.140 & 0.098 & 0.161 & 0.269 & 0.350 & 0.005 & 0.430 \\ 
\hdashline[1pt/1pt]
GSKET & MedIA’22 & 60$^{\spadesuit}$ & - & 0.496 & \textbf{0.327} & \textbf{0.238} & \textbf{0.178} & - & \underline{0.381} & - & - & - \\
CMCA & COLING’22 & -$^{\spadesuit}$ & - & 0.497 & 0.349 & 0.268 & 0.215 & 0.209 & 0.392 & - & - & - \\
M2KT & MedIA’23 & 60$^{\spadesuit}$ & - & \textbf{0.497} & 0.319 & 0.230 & 0.174 & - & \textbf{0.399} & - & - & - \\
DCL & CVPR’23 & 90$^{\spadesuit}$ & - & - & - & - & 0.163 & 0.193 & \textbf{0.383} & - & - & - \\
METransformer & CVPR’23 & -$^{\spadesuit}$ & - & 0.483 & 0.322 & 0.228 & 0.172 & 0.192 & 0.380 & - & - & - \\
MMTN & AAAI’23 & -$^{\spadesuit}$ & - & 0.486 & 0.321 & 0.232 & 0.175 & - & 0.375 & - & - & - \\ 
RAMT & TMM’24 & -$^{\spadesuit}$ & - & 0.482 & 0.310 & 0.221 & 0.165 & 0.195 & 0.377 & - & - & - \\ 
Pretrain+FMVP & TMM’24 & -$^{\spadesuit}$ & - & 0.485 & 0.315 & 0.225 & 0.169 & 0.201 & 0.398 & - & - & - \\ 
\midrule
\multirow{3}{*}{\textbf{FSE-0 (Ours)}} & \multirow{3}{*}{-} & 60 & 0.865 & 0.483 & 0.314 & 0.223 & 0.167 & 0.209 & 0.370 & \underline{0.375} & 0.222 & 0.578 \\
 &  & 90 & 0.865 & 0.475 & 0.308 & 0.219 & 0.164 & 0.206 & 0.369 & \textbf{0.375} & 0.219 & 0.576 \\
 &  & \textit{Cpl.} & 0.863 & 0.466 & 0.303 & 0.215 & 0.161 & 0.202 & 0.366 & 0.372 & 0.218 & 0.575 \\ 
\midrule
\multirow{3}{*}{\textbf{FSE-3 (Ours)}} & \multirow{3}{*}{-} & 60 & \textbf{0.869} & 0.495 & \underline{0.322} & 0.231 & 0.173 & \textbf{0.218} & 0.378 & \underline{0.375} & \underline{0.340} & \underline{0.604} \\
 &  & 90 & \textbf{0.869} & \textbf{0.498} & \textbf{0.324} & \textbf{0.232} & \textbf{0.174} & \textbf{0.216} & 0.378 & \textbf{0.375} & \underline{0.337} & \underline{0.602} \\
 &  & \textit{Cpl.} & \textbf{0.867} & \textbf{0.495} & \textbf{0.322} & \textbf{0.230} & \textbf{0.173} & \textbf{0.211} & \underline{0.375} & \underline{0.373} & \underline{0.335} & \underline{0.601} \\ 
\midrule
\multirow{3}{*}{\textbf{FSE-20 (Ours)}} & \multirow{3}{*}{-} & 60 & \underline{0.868} & \textbf{0.504} & \textbf{0.327} & \underline{0.234} & 0.174 & \underline{0.214} & 0.380 & 0.374 & \textbf{0.367} & \textbf{0.607} \\
 &  & 90 & \underline{0.868} & \underline{0.496} & \underline{0.323} & \underline{0.231} & \underline{0.172} & \underline{0.211} & \underline{0.380} & \underline{0.373} & \textbf{0.364} & \textbf{0.606} \\
 &  & \textit{Cpl.} & \textbf{0.867} & \underline{0.487} & \underline{0.317} & \underline{0.227} & \underline{0.168} & \underline{0.206} & \textbf{0.377} & 0.371 & \textbf{0.362} & \textbf{0.604} \\
\bottomrule
\end{tabular}
\end{table*}

\section{Experiments}
\subsection{Datasets, Evaluation Metrics, and Experimental Settings}

\textbf{Datasets.} We evaluate the effectiveness of our proposed FSE on the widely used benchmarks, MIMIC-CXR \cite{johnson-mimic-cxr-jpg} and IU X-ray \cite{demner2016preparing}. Consistent with previous works \cite{chen-etal-2020-generating, li-dcl, yang-gsket, tanida-rgrg}, we treat the FINDINGS section of radiology reports as reference reports. Regarding data partitioning, we follow the official settings for MIMIC-CXR, while for IU X-ray\footnote{Due to the absence of a standardized data partitioning for IU X-ray, experiments on this dataset are less convincing than those conducted on MIMIC-CXR.}, we adhere to the settings used in R2Gen \cite{chen-etal-2020-generating} and R2GenCMN \cite{chen-etal-2021-cross-modal}. Blank or clinically insignificant reference reports, such as entries containing only “\emph{Portable supine chest radiograph\_\_at 23:16 is subnitted.}”, are removed. After filtering, on the MIMIC-CXR dataset, the training set, validation set, and test set include 269,239 (150,957), 2,113 (1,182), and 3,852 (2,343) radiographs (reports), respectively. IU X-ray dataset comprises 4,136 (2,068), 592 (296), and 1,180 (590) radiographs (reports), respectively. All reproducibility methods utilize the same test set for fair comparison.

\textbf{Evaluation metrics.} To ensure a comprehensive evaluation, we employ conventional natural language generation (NLG) metrics and clinical efficacy (CE) metrics, which assess lexical similarity and clinical consistency between generated and reference reports. NLG metrics comprise BERTScore (BS, pre-trained on “\textit{distilbert-base-uncased}” \cite{Sanh2019DistilBERTAD}), BLEU-\textit{n} (BL-\textit{n}, where \(n\in \left \{ 1,2,3,4 \right \} \)), METEOR (MTR), and ROUGE-L (R-L). CE metrics comprises F\textsubscript{1,mic-14} CheXbert (CX14), F\textsubscript{1,mic-5} CheXbert (CX5) \cite{Smit2020_chexbert}, and F\textsubscript{1} RadGraph (RG) \cite{delbrouck-etal-2022-improving}. CX14 and CX5 metrics represent micro-F\textsubscript{1} scores based on 14 and 5 categories related to thoracic diseases and support devices, as labeled by CheXpert \cite{irvin-chexpert}. The RG metric \cite{yu2023evaluating} measures the overlap of clinical entities and relationships extracted by RadGraph, with a partial reward level. Higher values across all metrics indicate superior performance. CX5, CX14, and RG metrics are computed using the f1chexbert \cite{Smit2020_chexbert} and radgraph \cite{jain-radgraph} libraries, respectively. 

\begin{table*}
\centering
\caption{Ablation studies on MIMIC-CXR. In Stage 1, “F/R” denotes whether Factual serialization or the complete Report is used for alignment. In Stage 2, “F/R” indicates the presentation form of similar historical cases (Factual serialization/Report), while “Fusion” refers to the fusion method (CMF/Concatenation).}
\label{table: component}
\resizebox{\linewidth}{!}{
\begin{tabular}{ccccccccccc} 
\toprule
\multirow{2}{*}{\textbf{Model}} & \multicolumn{3}{c}{\textbf{Stage 1}} & \multicolumn{2}{c}{\textbf{Stage 2}} & \multicolumn{2}{c}{\textbf{NLG}$\uparrow$} & \multicolumn{3}{c}{\textbf{CE}$\uparrow$} \\ 
\cmidrule(lr){2-4}\cmidrule(lr){5-6}\cmidrule(r){7-8}\cmidrule(r){9-11}
 & \textbf{F/R} & ${\cal L}_{global}$ & ${\cal L}_{local}$ & \textbf{F/R} & \textbf{Fusion} & \textbf{BS} & \textbf{BL-2} & \textbf{RG} & \textbf{CX5} & \textbf{CX14} \\ 
\midrule
BASE & - & \ding{55} & \ding{55} & - & - & 0.841 & 0.209 & 0.211 & 0.339 & 0.338 \\
(a) & R & \ding{51} & \ding{51} & R & CMF & 0.837 & 0.199 & 0.184 & 0.377 & 0.305 \\
(b) & R & \ding{51} & \ding{51} & F & CMF & 0.838 & 0.198 & 0.182 & 0.342 & 0.292 \\
(c) & F & \ding{51} & \ding{55} & F & CMF & 0.842 & 0.199 & 0.207 & 0.453 & 0.379 \\
(d) & F & \ding{55} & \ding{51} & F & CMF & 0.836 & 0.187 & 0.185 & 0.347 & 0.294 \\
(e) & F & \ding{51} & \ding{51} & - & - & \textbf{0.845} & 0.206 & \underline{0.234} & \underline{0.513} & 0.431 \\
(f) & F & \ding{51} & \ding{51} & F & Cat & \underline{0.844} & 0.206 & 0.230 & 0.489 & 0.406 \\
(g) & F & \ding{51} & \ding{51} & R & CMF & 0.843 & \underline{0.216} & 0.233 & 0.512 & \underline{0.432} \\ 
\midrule
\textbf{FSE-5 (Ours)} & F & \ding{51} & \ding{51} & F & CMF & \textbf{0.845} & \textbf{0.227} & \textbf{0.236} & \textbf{0.560} & \textbf{0.482} \\
\bottomrule
\end{tabular}
}
\end{table*}

\textbf{Experimental settings.} The image and text encoder are detailed in Section \ref{sec: cross-modal alignment}, and the text decoder follows the memory-driven Transformer from R2Gen \cite{chen-etal-2020-generating}. For the MIMIC-CXR dataset, Stage 1 (factuality-guided contrastive learning) utilizes the AdamW optimizer with an initial learning rate of 5e-5, training for 100 epochs. In Stage 2 (evidence-driven report generation), the RAdam optimizer is employed with a learning rate of 5e-5 for 50 epochs. For the IU X-ray dataset, given its smaller size, we skip Stage 1 and proceed directly to Stage 2. The model is initialized with the pre-trained weights from MIMIC-CXR and trained with the RAdam optimizer at a learning rate of 1e-4 for 30 epochs. The optimal model is selected based on the cumulative scores from RG, CX14, and BL-4 on the validation set, and the final performance is evaluated on the test set.

\subsection{Main Results}
We evaluate chest X-ray report generation performance in both specific and general scenarios. Specifically, in the testing phase, we preserve the generated reports unaltered and truncate reference reports to a predetermined length \(M_{gt}\) to serve as ground truth. In the general scenario, \(M_{gt}\) refers to the full length of reference reports, labeled as “\textit{Cpl.}”. Otherwise, it represents a specific scenario with truncated reference reports. Our proposed FSE is compared against 14 generative-based approaches: \textbf{R2Gen} \cite{chen-etal-2020-generating}, \textbf{R2GenCMN} \cite{chen-etal-2021-cross-modal}, \textbf{GSKET} \cite{yang-gsket}, \textbf{CMCA} \cite{song-etal-2022-cross}, \textbf{CvT2DistillGPT2} \cite{nicolson-improving}, \textbf{M2KT} \cite{yang-m2kt}, \textbf{DCL} \cite{li-dcl}, \textbf{METransformer} \cite{wang2023metransformer}, \textbf{OpenLLaMA-7B} \cite{lu2023effectively}, \textbf{SA} \cite{yan2023style}, \textbf{MMTN} \cite{aaai-23-MMTN}, \textbf{CAMANet} \cite{2024-CAMANet-JBHI-compare}, \textbf{RAMT} \cite{RAMT-2024-TMM}, and \textbf{FMVP} \cite{FMVP-2024-TMM} as well as a retrieval-based approach: \textbf{X-REM} \cite{pmlr-v227-jeong24a-MIDL-retrieve}. Notably, CXR-RePaiR \cite{pmlr-v158-endo21a-retrieve-ML4H} focuses on generating IMPRESSION sections, and RGRG \cite{tanida-rgrg} follows a different data partitioning strategy. As a result, these two methods are excluded from our comparisons. The comparative results for the MIMIC-CXR and IU X-ray datasets are presented in Tables \ref{table: main results} and \ref{table: main-results-iu}, respectively. FSE-\textit{n} refers to our proposed FSE that generates reports by referencing \textit{n} similar historical cases. Due to the absence of $\boldsymbol{M}_{gt}$ values in certain methods, their results are included for reference purposes only.

As evident from Table \ref{table: main results}, FSE-1 demonstrates competitive performance compared to both retrieval-based and generative-based approaches, with generative-based methods generally outperforming retrieval-based ones. Specifically, FSE-1 achieves either the best or second-best performance in both NLG and CE metrics across specific scenarios (i.e., \({M_{gt}} \in \left\{ {60,80,90,100} \right\}\)) and general scenarios (i.e., \(M_{gt}=Cpl.\)). Notably, with \(M_{gt}=80\), FSE-1 reaches scores of 0.397 for BL-1, 0.238 for RG, and 0.470 for CX14. Although FSE-1 slightly trails behind CvTDistillGPT2 \cite{nicolson-improving} in some NLG metrics at \(M_{gt}=60\), it shows an overall performance improvement of 4.1\%. As presented in Table \ref{table: main-results-iu}, our FSE-20 exhibits a slight advantage over peer methods, particularly excelling in the CX5 and CX14 metrics. Specifically, FSE-20 achieves scores of 0.504 for BL-1 and 0.607 for CX14, highlighting its effectiveness in clinical consistency. 

\begin{table*}
\centering
\caption{Variation in FSE performance with the number of similar historical cases on MIMIC-CXR. For each \({\boldsymbol{M_{gt}}}\), the best and second-best values are highlighted in \textbf{bold} and \underline{underlined}, respectively.}
\label{table: num cases}
\resizebox{\linewidth}{!}{
\begin{tabular}{llcccccccccc} 
\toprule
\multirow{2}{*}{\textbf{Method}} & \multirow{2}{*}{\({\boldsymbol{M_{gt}}}\)} & \multicolumn{7}{c}{\textbf{NLG}$\uparrow$} & \multicolumn{3}{c}{\textbf{CE}$\uparrow$} \\ 
\cmidrule(lr){3-9}\cmidrule(lr){10-12}
 &  & \textbf{BS} & \textbf{BL-1} & \textbf{BL-2} & \textbf{BL-3} & \textbf{BL-4} & \textbf{MTR} & \textbf{R-L} & \textbf{RG} & \textbf{CX5} & \textbf{CX14} \\ 
\midrule
\multirow{2}{*}{FSE-0} & 100 & \textbf{0.845} & 0.349 & 0.213 & 0.143 & 0.102 & 0.142 & 0.278 & 0.233 & 0.512 & 0.430 \\
 & \textit{Cpl.} & \textbf{0.845} & 0.338 & 0.206 & 0.137 & 0.098 & 0.138 & 0.277 & 0.234 & 0.513 & 0.431 \\ 
\hdashline[1pt/1pt]
\multirow{2}{*}{FSE-1} & 100 & \textbf{0.845} & \textbf{0.384} & \textbf{0.235} & \textbf{0.158} & \textbf{0.113} & \textbf{0.152} & \textbf{0.281} & \textbf{0.241} & 0.538 & \underline{0.473} \\
 & \textit{Cpl.} & \textbf{0.845} & \textbf{0.373} & \textbf{0.228} & \textbf{0.152} & \textbf{0.109} & \textbf{0.148} & \textbf{0.279} & \textbf{0.241} & 0.542 & 0.474 \\ 
\hdashline[1pt/1pt]
\multirow{2}{*}{FSE-3} & 100 & 0.843 & 0.369 & 0.223 & 0.148 & 0.105 & 0.146 & 0.276 & 0.230 & 0.541 & 0.458 \\
 & \textit{Cpl.} & 0.843 & 0.358 & 0.216 & 0.144 & 0.102 & 0.143 & 0.275 & 0.231 & 0.545 & 0.459 \\ 
\hdashline[1pt/1pt]
\multirow{2}{*}{FSE-5} & 100 & \underline{0.844} & \textbf{0.384} & \underline{0.234} & \underline{0.157} & \underline{0.112} & \underline{0.151} & \underline{0.279} & 0.235 & \textbf{0.556} & \textbf{0.481} \\
 & \textit{Cpl.} & \textbf{0.845} & \textbf{0.373} & \underline{0.227} & \underline{0.151} & \underline{0.108} & \textbf{0.148} & \underline{0.278} & 0.236 & \textbf{0.560} & \textbf{0.482} \\ 
\hdashline[1pt/1pt]
\multirow{2}{*}{FSE-10} & 100 & \underline{0.844} & \underline{0.379} & 0.232 & 0.156 & 0.111 & 0.151 & \underline{0.279} & \underline{0.236} & \underline{0.548} & \underline{0.473} \\
 & \textit{Cpl.} & \underline{0.844} & \underline{0.368} & 0.224 & 0.150 & 0.107 & 0.147 & \underline{0.278} & \underline{0.237} & \underline{0.550} & \underline{0.475} \\ 
\hdashline[1pt/1pt]
\multirow{2}{*}{FSE-20} & 100 & 0.843 & 0.347 & 0.210 & 0.139 & 0.099 & 0.141 & 0.273 & 0.229 & 0.537 & 0.455 \\
 & \textit{Cpl.} & 0.843 & 0.336 & 0.202 & 0.133 & 0.094 & 0.137 & 0.272 & 0.230 & 0.541 & 0.457 \\
\bottomrule
\end{tabular}
}
\end{table*}

\begin{figure*}
    \centering
    \includegraphics[width=1\linewidth]{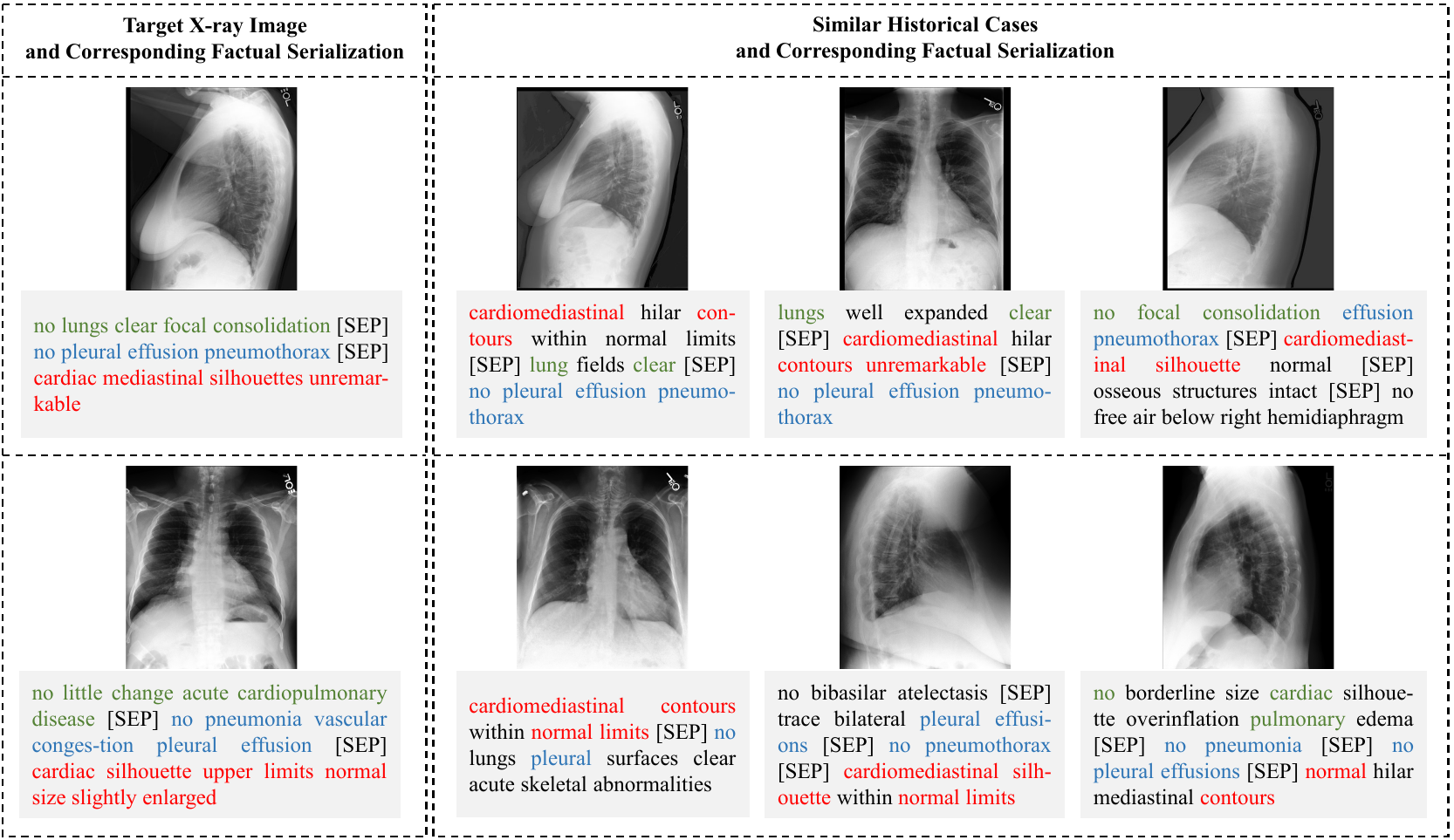}
    \caption{Visualization of similar historical cases for two samples from the MIMIC-CXR test set. In the left panel, factual sequences in the target X-ray images are shown in different colors. The right panel highlights corresponding factual sequences in similar historical cases using colors that match those in the target X-ray images.}
    \label{fig:4}
\end{figure*}

\subsection{Ablation Study}
\textbf{Contribution of each component.} Ablation studies on the MIMIC-CXR dataset were conducted to assess the contribution of each component, as shown in Table \ref{table: component}. “BASE” refers to R2Gen \cite{chen-etal-2020-generating}, which utilizes the same image encoder and text decoder as our FSE method. When similar historical cases are injected in Stage 2, the number is fixed at 5 per sample. If the “F/R” in Stage 1 and Stage 2 are identical, the text encoder is frozen during training. Experimental results demonstrate the positive effect of each component:
\begin{itemize}
    \item “BASE” outperforms (a) and (b), suggesting that using complete reports for cross-modal alignment is challenging, possibly due to interference from the presentation-style vocabulary.
    \item Compared to “BASE”, (e) exhibits a notable enhancement in CE metrics, highlighting the benefit of factuality-guided contrastive learning in enhancing the clinical effectiveness of generated reports.
    \item FSE-5 exceeds (c) and (d), indicating that effective alignment requires both \({\cal L}_{global}\) and \({\cal L}_{local}\) losses.
    \item Compared to (g) and FSE-5, (f) performs slightly worse in both NLG and CE metrics, implying that the cross-modal fusion network (CMF) better captures relevant empirical features from similar historical cases than simple concatenation, thereby assisting the text decoder in producing high-quality reports.
    \item FSE-5 outperforms (e) by 2.1\%, 4.7\%, and 5.1\% for BL-2, CX5, and CX14, respectively, suggesting that similar historical cases are crucial for generating clinically accurate reports by providing empirical insights that guide the text decoder. 
    \item FSE significantly improves NLG and CE metrics over (a), (b), and (g), highlighting the critical role of factual serialization in both factuality-guided contrastive learning and evidence-driven report generation.
\end{itemize}

\begin{figure*}
    \centering
    \includegraphics[width=1\linewidth]{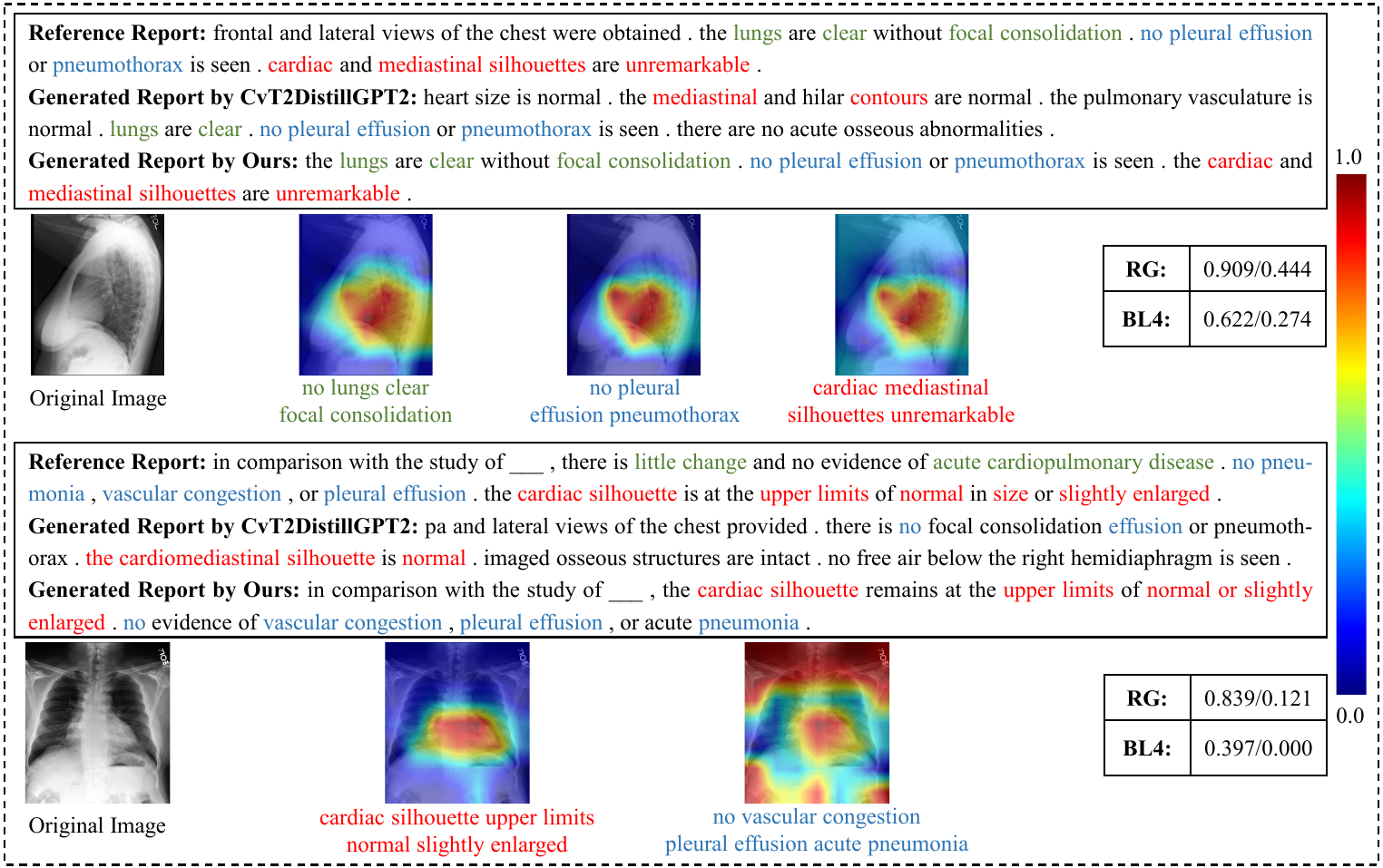}
    \caption{Two examples of generated reports and attention visualization in the MIMIC-CXR test set. In the reference report, different colors are used to highlight factual sequences, with matching colors applied to corresponding factual vocabulary in the generated report. Performance metrics, represented as "A/B", indicate the values achieved by FSE-5 and CvT2DistillGPT2 \cite{nicolson-improving}, respectively.}
    \label{fig:5}
\end{figure*}

\textbf{Impact of the number of similar historical cases.} To investigate how the number of similar historical cases affects performance, we train the FSE model on MIMIC-CXR with quantities ranging from \(\left \{ 0,1,3,5,10, 20 \right \} \). Performance is evaluated from both specific and general scenario perspectives, as detailed in Table \ref{table: num cases}. Integrating similar historical cases offers relevant empirical insights that guide the text decoder in producing clinically accurate reports, resulting in significant improvements in both NLG and CE metrics. Specifically, in the specific scenario (i.e., \(M_{gt}=100\)), FSE-1 improves BL-1 by 3.5\% and CX14 by 4.3\% compared to FSE-0. Similar benefits are also observed in the general scenario, demonstrating the benefits of incorporating similar historical cases. While FSE-1 excels in conventional NLG metrics and the RG metric, FSE-5 shows the most balanced performance across NLG and CE metrics. However, a decrease in performance is noted when the number of cases exceeds 5, likely due to increased model complexity and potential overfitting.


\subsection{Qualitative Analysis}
\textbf{Visualization of similar historical cases.} To illustrate the effectiveness of similar historical cases, we visualize three such cases for two samples from the MIMIC-CXR test set, as shown in Fig. \ref{fig:4}. A closer color distribution between similar historical cases and the target X-ray image indicates more comprehensive clinical coverage, while longer color bars reflect more detailed findings. We observe that factual serialization in similar historical cases aligns with the target X-ray image, thereby offering valuable empirical insights for the text decoder. Excitingly, even with significant differences in the view positions between the target X-ray image and similar historical cases, we successfully retrieve cases with similar factual sequences, without relying on disease labels. This highlights the effectiveness of factuality-guided contrastive learning in extracting aligned visual features for retrieving relevant historical cases.

\textbf{Visualization of generated reports.} To illustrate our generated reports, we present two examples from the MIMIC-CXR test set in Fig. \ref{fig:5}. A closer color distribution between generated and reference reports signifies more comprehensive clinical coverage, while longer color bars reflect more detailed findings. Using the multi-head attention mechanism in the decoder's first layer, we visualized the attention weights assigned to factual sequences in the generated reports. The results reveal that FSE-5 outperforms the baseline in both matched colors and performance metrics. Additionally, our model exhibits a degree of interpretability by localizing factual sequences within the original image.

\section{Conclusion}
In this paper, we have proposed Factual Serialization Enhancement (FSE), a two-stage method for generating chest X-ray reports. First, we presented factuality-guided contrastive learning to align chest X-ray images with factual serialization derived from our proposed structural entities approach, alleviating the impact of presentation-style vocabulary on cross-modal alignment. Subsequently, we introduced evidence-driven report generation, enhancing diagnostic accuracy by integrating insights from similar historical cases. Experiments on MIMIC-CXR and IU X-ray datasets demonstrated that factual serialization plays critical roles in factuality-guided contrastive learning and evidence-driven report generation. In addition, integrating similar historical cases offers relevant empirical insights that guide the text decoder in producing clinically accurate reports, resulting in significant improvements in both NLG and CE metrics. In the future, we will explore modeling temporal information \cite{recap} to capture disease progression and predicting uncertainty \cite{uncertainty} to further enhance model reliability.

\bibliographystyle{IEEEtran}
\bibliography{refs}
\end{document}